\documentclass[final,5p,times]{elsarticle}
\usepackage{amssymb,bbold}
\usepackage{natbib}
\usepackage{hyperref}
\usepackage[latin1]{inputenc}
\usepackage{float}
\usepackage{graphicx}
\usepackage{graphics}
\usepackage{amsfonts}
\usepackage{color}
\usepackage[T1]{fontenc}
\usepackage[english]{babel}
\usepackage{wrapfig}
\usepackage{amsmath}
\usepackage{amssymb}
\usepackage{mathrsfs}

\hypersetup{
  colorlinks=true,linkcolor=blue,citecolor=blue,
  filecolor=blue,urlcolor=blue,breaklinks=true
}
\biboptions{sort&compress}
\journal{Physics Letters A}
\begin{document}

\begin{frontmatter}

\title{2DEG on a cylindrical shell with a screw dislocation}

\author[ufcg]{Cleverson Filgueiras}
\ead{cleversonfilgueiras@yahoo.com.br}
\author[ufma]{Edilberto O. Silva}
\ead{edilbertoos@pq.cnpq.br}

\address[ufcg]{Unidade Acadêmica de F\'{i}sic,
  Universidade Federal de Campina Grande, POB
  10071, 58109-970, Campina Grande, Para\'{i}ba, Brazil}

\address[ufma]{Departamento de F\'{i}sica,
  Universidade Federal do Maranh\~{a}o,
  Campus Universit\'{a}rio do Bacanga,
  65085-580, S\~{a}o Lu\'{i}s, Maranh\~{a}o, Brazil}

\begin{abstract}
A two dimensional electron gas on a cylindrical surface with a screw dislocation is considered. More precisely, we investigate how both the geometry and the deformed potential due to a lattice distortion affect the Landau levels of such system. The case showing the deformed potential can be thought in the context of 3D common semiconductors where the electrons are confined on a cylindrical shell. We will show that important quantitative differences exist due to this lattice distortion. For instance, the effective cyclotron frequency is diminished by the deformed potential, which in turn enhances the Hall conductivity.
\end{abstract}


\end{frontmatter}

\section{Introduction}
\label{intro}

The influence of screw dislocations in quantum systems has received
considerable attention over the last years. Some works are based on the
geometric theory of defects in semiconductors developed by Katanaev and Volovich \cite{katanaev}. In this approach, the semiconductor with
a screw dislocation is described by a Riemann-Cartan manifold where the
torsion is associated to the Burgers vector. In this continuum limit, a
screw dislocation affects a quantum system like an isolated magnetic flux
tube, causing an Aharonov-Bohm (AB) interference phenomena \cite{ab}%
. The energy spectrum of electrons around this kind of defect shows a
profile similar to that of the AB system \cite{e1,e2,e3,moraes,knut}%
. These works describe the effect due to the geometric electron motion. A
second ingredient, which may show pronounced influences in these quantum
systems, is an additional \textit{deformed potential} induce by a lattice
distortion \cite{distorted}. It is a repulsive scalar
potential(noncovariant). The impact of this potential was first investigated
in Ref. \cite{impact}, where the scattering of electrons around a
screw dislocation was investigated. Recently, it was showed that a single
screw dislocation has profound influences on the electronic transport
in semiconductors \cite{hisao}. Both contributions, the covariant
and noncovariant terms, were taken into account. Inspired in these works, in
this paper, we will investigate how the deformed potential due to a lattice
distortion affects the energy levels of a two dimensional electron gas
(2DEG) confined on a cylindrical surface in a 3D semiconductor. The case
with the absence of such potential described in the literature can find
applications in the context of carbon nanotubes \cite%
{lorenci,anibal,breno}. We will show that, for a 2DEG confined on a
cylindrical shell in a 3D semiconductor like silicon, important quantitative
differences exist due to this noncovariant term. The procedure to confine
electrons on a curved surface is also considered, given rise to a geometric
potential induced by such confinement \cite{costa}.

This work is divided as follows. In Sec. 2, we derive the Schr\"{o}dinger equation for a 2DEG on a cylindrical surface with the elastic deformation induced by the screw deslocation.
In Sec. 3, we investigate how the deformed potential influences the
energy levels of a 2DEG in the presence of an external magnetic field on
such cylinder. This case can find applications in the context of quantum
Hall effects, for instance. Next, we have the concluding remarks.

\section{The Schrodinger equation for a 2DEG on the deformed cylinder}

\label{sec:1}

In this section, we derive the Schrodinger equation for the 2DEG on a
cylinder with a deformation induced by a screw dislocation. We consider an
infinitely long linear screw dislocation oriented along the $z$%
-axis. The three-dimensional geometry of the medium is
characterized by a torsion which is identified with the surface density of
the Burgers vector in the classical theory of elasticity. The metric of the
medium with this kind of defect is given (in cylindrical coordinates) by \cite{katanaev}
\begin{equation}
ds^{2}=\left( dz+\beta d\phi \right) ^{2}+d\rho ^{2}+\rho ^{2}d\phi ^{2},
\label{3dmetric}
\end{equation}%
In Eq. (\ref{3dmetric}), $\left( \rho
,\phi ,z\right) \rightarrow \left( \rho ,\phi +2\pi ,z\right) $ and $\beta $
is a parameter related to the Burgers vector $b$ by $\beta =b/2\pi $. The
induced metric describes a flat medium with a singularity at the origin. The
only non-zero component of the torsion tensor in this case is given by the
two form
\begin{equation}
T^{1}=2\pi \beta \delta ^{2}(\rho )d\rho \wedge d\phi ,  \label{curv}
\end{equation}%
with $\delta ^{2}(\rho )$ being the two-dimensional delta function in the
flat space. Figure 1 illustrates the formation of a screw dislocation in the
bulk of a 3D crystal.
\begin{figure}[tbh]
\begin{center}
\includegraphics[height=3cm]{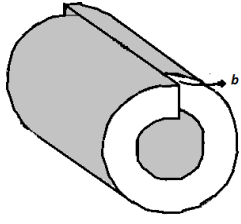}\label{f5}
\end{center}
\caption{{\protect\scriptsize Cylindrical portion of a 3D solid showing the
dislocation.}}
\label{eg}
\end{figure}
We will consider an electron gas confined just on a cylindrical shell.
Putting $\rho \equiv R=\mathrm{constant}$, the two dimensional line element
will be given by
\begin{equation}
dl^{2}=\left( dz+\beta d\phi \right) ^{2}+R^{2}d\phi ^{2},  \label{2dmetric}
\end{equation}

In Ref. \cite{lorenci}, it was showed that the geometry of a
cylindrical shell with a screw dislocation is equivalent to that of a
cylindrical shell without a screw dislocation but with radius $r=\left(
R^{2}+\beta ^{2}\right) ^{1/2}$. This way, the line element \ref{2dmetric}
is rewritten as
\begin{equation}
dl^{2}=dZ^{2}+r^{2}d\theta ^{2},  \label{2dmetric2}
\end{equation}%
where
\begin{equation}
\theta =\phi +\frac{\beta Z}{r\sqrt{r^{2}-\beta ^{2}}}, \;\;
Z =\frac{z}{r}\sqrt{r^{2}-\beta ^{2}}.  \label{def}
\end{equation}%
This metric obeys the same usual identification, $\left( \theta ,Z\right)
\rightarrow \left( \theta +2\pi ,Z\right) $.

Unlike Ref. \cite{lorenci}, which considers possible applications
in the context of carbon nanotubes, we will consider common semiconductors.
This way, we have to introduce a \textit{deformed potential} which describes
the effects of the lattice deformation on the electronic properties in such
materials \cite{distorted}. For a screw dislocation, it is found to be
\begin{equation}
V_{d}(\rho )=\frac{\hbar ^{2}}{2ma^{2}}\frac{b^{2}}{4\pi ^{2}\rho ^{2}}\left[
2+a^{2}\left( \frac{\partial }{\partial z}\right) ^{2}\right] ,
\label{deformedV}
\end{equation}%
where $a$ is the lattice constant.

We now consider the existence of a second quantum potential. The procedure
to confine an electron gas on a surface is based on the Da Costa's
approach \cite{costa}: we consider the charge carriers on a thin interface
which is a infinity non flat quantum well. At the end, we consider the
thickness of such interface going to zero and then we separate the
transverse Schrodinger equation from the longitudinal one. In the transverse
direction, the electrons are frozen due to the infinity quantum well, while
they are free along the surface. The consequence is that a geometric
potential arises. For electrons on a cylindrical surface, it is
given by \cite{costa}
\begin{equation}
V_{g}(R)=-\frac{\hbar ^{2}}{8mR^{2}}.  \label{dacostav}
\end{equation}

The Schrodinger equation for a quantum particle in a background $g_{\mu \nu
} $ in the presence of both potentials described above is given by
\begin{equation}
-\frac{\hbar ^{2}}{2m}\frac{1}{\sqrt{g}}\partial _{\mu }\sqrt{g}g^{\mu \nu
}\partial _{\nu }\Psi +\left[ V_{d}(R)+V_{g}(R)\right] \Psi  =E\Psi ,
\end{equation}%
where $g\equiv \det {g_{\mu \nu }}$. For a cylindrical shell with the metric %
\ref{2dmetric2}, we have
\begin{equation}
-\frac{\hbar ^{2}}{2m}\left[ \frac{1}{r^{2}}\frac{\partial ^{2}}{\partial
\theta ^{2}}+\frac{\partial ^{2}}{\partial Z^{2}}\right] \Psi +\left[
V_{d}(R)+V_{g}(r)\right] \Psi=E\Psi .
\end{equation}%

In the next section, we will consider the existence of a constant external magnetic field  perpendicular to the cylinder height.

\section{Landau levels of a 2DEG on a cylinder with screw dislocation}

In this section, we investigate the influence of the deformed
potential (\ref{deformedV}) on the energy levels of a 2DEG on a cylinder
with a screw dislocation in common semiconductors. Since only the component
of the magnetic field pointing along the surface normal governs the Lorentz
force and the electronic transport in 2DEGs, we consider the presence
of a uniform magnetic field crossing the cylinder transversely. Then, the
vector potential in the symmetric gauge is
\begin{equation}
A_{Z}=Br\sin \theta .  \label{gauge}
\end{equation}%
This way, the Schrödinger equation in the background of Eq.
(\ref{2dmetric2}) with such external magnetic field is \cite%
{ferrari}
\begin{equation}
-\frac{\hbar ^{2}}{2mr^{2}}\frac{\partial ^{2}\Psi }{\partial
\theta ^{2}}+\frac{1}{2m}\left[ -i\hbar \frac{\partial }{\partial Z}%
-eA_{Z}\right] ^{2}\Psi +\left[ V_{d}(R)+V_{g}(r)\right] \Psi =E\Psi .  \label{schocyla}
\end{equation}%
Since (\ref{deformedV}) depends on the linear momentum operator in the $z$%
-direction, we also considered the minimal coupling in it, $p_{z}\rightarrow
p_{Z}-eA_{Z}$, together with the relation (\ref{def}). Considering $\Psi
=e^{ik_{Z}Z}\psi (\theta )$, the Schrödinger equation to be studied
is
\begin{equation}
-\frac{\hbar ^{2}}{2mr^{2}}\frac{\partial ^{2}\psi }{\partial
\theta ^{2}}+\frac{m\omega _{c}^{2}}{2}\left( 1+\frac{\beta ^{2}}{%
R^{2}}\right) ^{-1}\left[ r\sin \theta -\frac{\hbar k_{Z}}{eB}\right]
^{2}\psi =\epsilon \psi ,  \label{schocyl}
\end{equation}%
where $\epsilon =E+\frac{\hbar ^{2}}{8mR^{2}}\left( 1+\frac{\beta ^{2}}{R^{2}%
}\right) ^{-1}-\frac{\hbar ^{2}\beta ^{2}}{ma^{2}R^{2}}$. By
defining the coordinate $x\equiv \theta /r$, we can rewrite Eq. (%
\ref{schocyl}) as
\begin{equation}
-\frac{\hbar ^{2}}{2m}\frac{\partial ^{2}\psi }{\partial x^{2}}+%
\frac{m\omega _{c}^{2}}{2}\left( 1+\frac{\beta ^{2}}{R^{2}}\right)
^{-1}\left[ r\sin \left( x/r\right) -\frac{\hbar k_{Z}}{eB}\right] ^{2}\psi=\epsilon \psi .  \label{parabola}
\end{equation}%
Considering the approximation for a sufficiently strong normal to the
surface component of the magnetic field and for electrons with the Landau
oscillator suspension center far enough from the side edges of a cylindrical
strip, we expand the potential present in Eq. (\ref{parabola})
around the minimum $x_{M}=r\sin ^{-1}\left( \hbar K_{Z}/eBr\right) $ up to
the harmonic term. The result of this operation provides us the
following eigenvalue equation:
\begin{equation}
-\frac{\hbar ^{2}}{2m}\frac{\partial ^{2}\psi }{\partial x^{2}}+%
\frac{m\omega ^{2}}{2}\left[ x-x_{M}\right] ^{2}\psi =\epsilon \psi .
\label{harmonic}
\end{equation}%
with the effective cyclotron frequency
\begin{eqnarray}
\omega  &\equiv &\omega _{c}\left( 1+\frac{\beta ^{2}}{R^{2}}\right) ^{-%
\frac{1}{2}}\cos \left[ \sin ^{-1}\left( \frac{\hbar k_{Z}}{eBr}\right) %
\right] ,  \notag \\
&=&\omega _{c}\left( 1+\frac{\beta ^{2}}{R^{2}}\right) ^{-\frac{1}{2}}\sqrt{%
1-\left( \frac{\hbar k_{Z}}{eBr}\right) ^{2}}.
\end{eqnarray}%
Notice that $\omega <\omega _{c}$.

\begin{figure}[h!]
\begin{center}
\includegraphics[height=7cm]{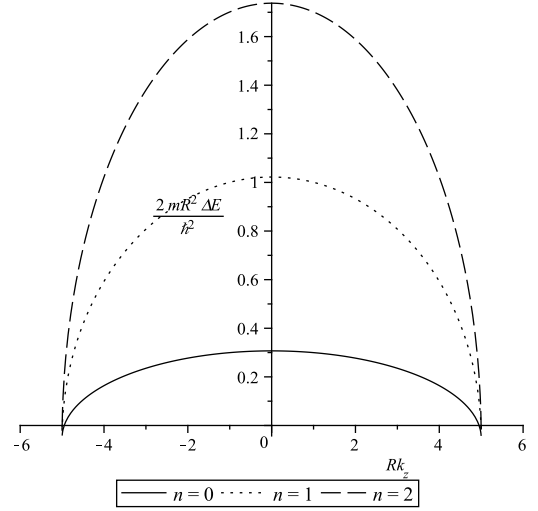}
\end{center}
\vspace{1.0cm}
\caption{{\protect\scriptsize The energy difference, $\Delta \equiv
\left\vert E_{n}^{\prime}-E_{n}\right\vert $, between two models: one with
and the other without the deformed potential. $\Delta $ versus $%
Rk_{z}$ for $\protect\alpha =10$ and $\protect\lambda =0.4$. Notice the
dispersion curves with maximum value at $k_{z}=0$.}}
\label{fig2a}
\end{figure}

\begin{figure}[h!]
\begin{center}
\includegraphics[height=7cm]{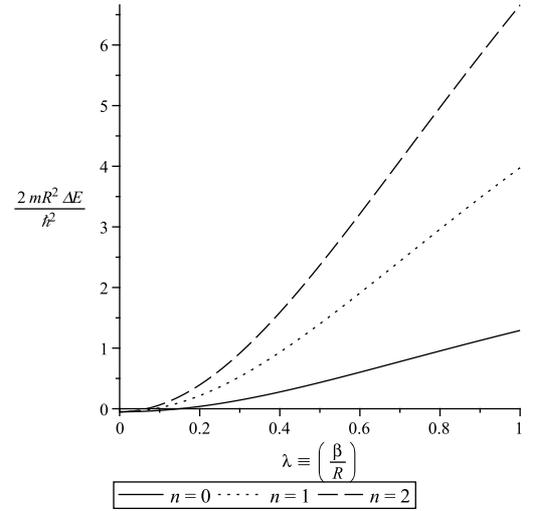}
\end{center}
\vspace{1.0cm}
\caption{{\protect\scriptsize The energy difference, $\Delta \equiv
\left\vert E_{n}^{\prime}-E_{n}\right\vert $, between two models: one with
and the other without the deformed potential. $\Delta $
versus the ratio between the Burges vector and the radius of the cylinder, $%
\protect\beta/R=1$($\protect\alpha =10$ and $Rkz=2$). The energy difference
is increased until $\protect\beta/R=1$.}}
\label{fig2b}
\end{figure}

\begin{figure}[h!]
\begin{center}
\includegraphics[height=7cm]{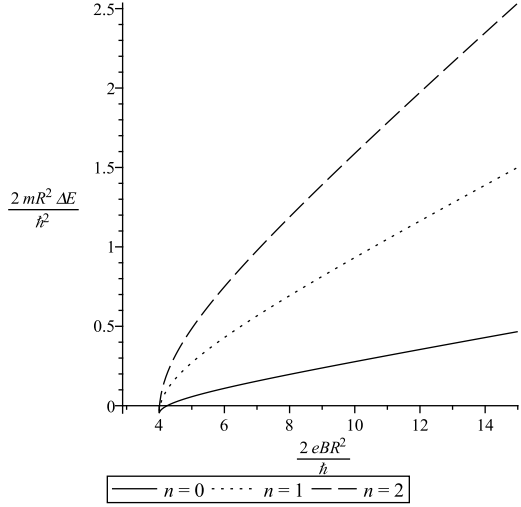}
\end{center}
\vspace{1.0cm}
\caption{{\protect\scriptsize The energy difference, $\Delta \equiv
\left\vert E_{n}^{\prime}-E_{n}\right\vert $, between two models: one with
and the other without the deformed potential. $\Delta $ versus $%
\protect\alpha \equiv (2eBR^{2})/\hbar$ for $Rk_{z}=2 $ and $\protect\lambda %
=0.4$. $\Delta $ is pronounced for higher magnetic fields. See that there is
one value of the magnetic field to which $\Delta =0$.}}
\label{fig2c}
\end{figure}

The eigenvalues of Eq. (\ref{harmonic}) are the Landau levels and
the energy levels are
\begin{eqnarray}
E_{n} &=&\left( n+\frac{1}{2}\right) \hbar \omega _{c}\left( 1+\frac{\beta
^{2}}{R^{2}}\right) ^{-\frac{1}{2}}\sqrt{1-\left( \frac{\hbar k_{Z}}{eBr}%
\right) ^{2}}  \notag  \\
&-&\frac{\hbar ^{2}}{8mR^{2}}\left( 1+\frac{\beta ^{2}}{R^{2}}\right) ^{-1}+%
\frac{\hbar ^{2}}{m}\frac{\beta ^{2}}{a^{2}R^{2}}. \label{landau}
\end{eqnarray}%
For $\beta \rightarrow 0$, we have \cite{iran}
\begin{equation*}
E_{n}=\left( n+\frac{1}{2}\right) \hbar \omega _{c}\sqrt{1-\left( \frac{%
\hbar k_{z}}{eBR}\right) ^{2}}-\frac{\hbar ^{2}}{8mR^{2}},
\end{equation*}%
while that for $R\rightarrow \infty $, we have the Landau levels for
electrons on a flat sample. Considering $b/a=1$\cite{tech} and defining the
dimensionless parameters $q\equiv rk_{Z}=Rk_{z}$, $\alpha \equiv
2eBR^{2}/\hbar $ and $\lambda ^{2}\equiv \beta ^{2}/R^{2}$, we put Eq.(\ref%
{landau}) in the following form:
\begin{eqnarray}
\frac{2mR^{2}E_{n}}{\hbar ^{2}}&=&\left( n+\frac{1}{2}\right) \alpha \left(
1+\lambda ^{2}\right) ^{-\frac{1}{2}}\sqrt{1-4\left( \frac{q^{2}}{\alpha ^{2}%
}\right) }  \notag \\ &-&\frac{1}{4}\left( 1+\lambda ^{2}\right) ^{-1}+\frac{1}{4\pi ^{2}}.
\label{dimensionless}
\end{eqnarray}%
In the absence of the deformed potential, the energy of the system
is \cite{breno}
\begin{equation}
\frac{2mR^{2}E_{n}^{\prime }}{\hbar ^{2}} =\left( n+\frac{1}{2}\right)
\alpha \sqrt{1-4\left( \frac{q^{2}}{\alpha ^{2}}\right) }-\frac{1}{4}\left( 1+\lambda ^{2}\right) ^{-1}.  \label{delta}
\end{equation}%
From this expression, we can see how the presence of a deformed
potential has a pronounced influence since when it is absent, the Landau
levels are just shifted by the geometric potential. In Figs. \ref{fig2a}, \ref{fig2b} and \ref{fig2c}, we evaluate the difference between these two
cases given by Eqs. (\ref{dimensionless}) and (\ref{delta}). We
plot the energy difference which is $\Delta =\left\vert E_{n}-E_{n}^{\prime
}\right\vert $. In Fig. \ref{fig2a}, we have the energy difference versus $%
Rk_{z}$. The maximum difference is at $k_{z}=0$.
This figure shows that, unlike the case of a 2DEG on a flat sample, the Landau levels are not
dispersionless. In Fig. \ref{fig2b}, we can see that the difference between the two
spectra is pronounced as the ratio between the Burgers vector and the radius
of the cylinder is increased. The energy difference versus the magnetic
field is provided in Fig. \ref{fig2c}. It can be seeing that it is increased as the
magnetic field intensity is increased. There is a value of the magnetic
field for which $\Delta =0$.
After examining the influence of the deformed potential due to lattice
distortions, we now analyze how the screw dislocation affects the energy
levels in comparison to the case of a flat sample.

\begin{figure}[h!]
\begin{center}
\includegraphics[height=7cm]{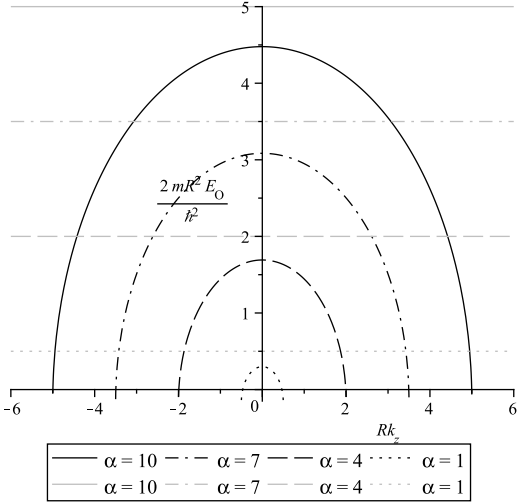}
\end{center}
\vspace{1.0cm}
\caption{{\protect\scriptsize $E_{n}$ versus $Rk_{z}$ for different values
of the magnetic field. The curves are also equally spaced and the maximum
values of energy are lower than the Landau levels in a flat sample.}}
\label{fig3}
\end{figure}

\begin{figure}[h!]
\begin{center}
\includegraphics[height=7cm]{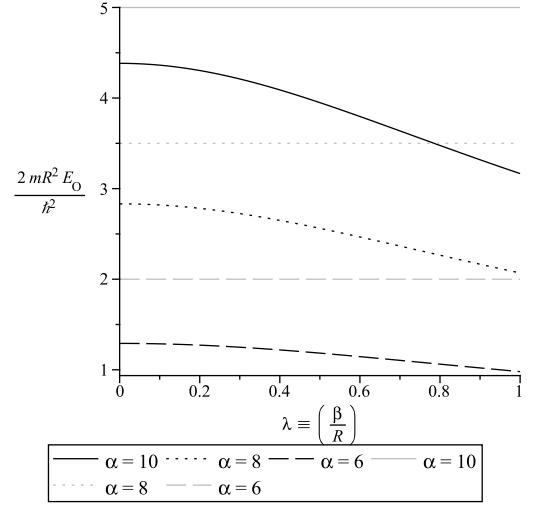}
\end{center}
\vspace{1.0cm}
\caption{{\protect\scriptsize Plot of $E_{0}$ versus $\frac{\protect\beta }{R%
}$(ratio between the Burger's vector and the cylinder radius), for different
values of the magnetic field.}}
\label{fig4}
\end{figure}

\begin{figure}[h!]
\begin{center}
\includegraphics[height=7cm]{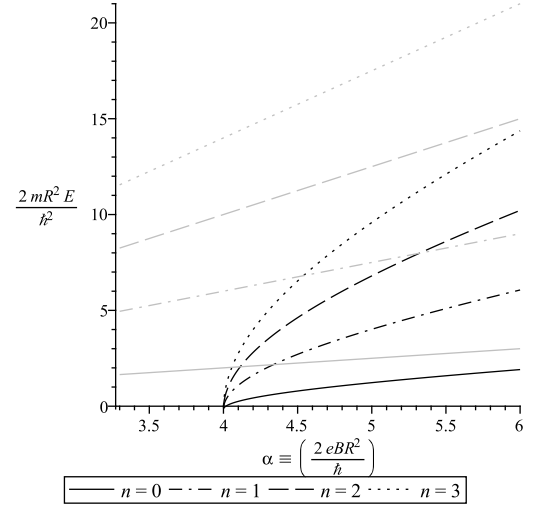}
\end{center}
\vspace{1.0cm}
\caption{{\protect\scriptsize Energy levels as function of $\protect\alpha \equiv \frac{2eB}{%
\hbar }R^{2}$ for $\protect\lambda =0.4$ and $Rk_{z}=2$. Due to dispersion,
these energies are no longer linear with respect the magnetic field.}}
\label{fig5}
\end{figure}

In Fig. \ref{fig3}, we plot the dispersion curves, that is, $E_{n}$
versus $Rk_{z}$ for different values of the magnetic field. The curves are
also equally spaced and the maximum values of energy are lower than the
Landau levels in a flat sample.
In Fig. \ref{fig4} we investigate how the screw dislocation affects
the ground state energy, $n=0$. We consider different values of the magnetic
field. Notice that, when $\beta /R=0$, there is still the geometric
potential (\ref{dacostav}) due to curvature and for this reason the energy
levels are lower than in the case of a flat sample.
Finally, in Fig. \ref{fig5}, the energy levels are depicted as
function of magnetic field. For a flat sample, it is well known that the
Landau levels are linear with respect the magnetic field intensity. The
dispersion due to curvature changes this fact.
In Ref. \cite{ballistic}, the energy spectrum and the ballistic
transport of a 2DEG on a cylindrical surface were theoretically
investigated. The authors investigated the hall conductivity in the case
where only the lowest Landau band is occupied (ultra-quantum limit),
the case when more than one \textit{non-overlapping} Landau bands are
occupied and the case when more than one \textit{overlapping} Landau bands
are occupied. In order to appreciate the influence of a screw dislocation on
such phenomenon, we consider just the ultra-quantum limit. To see the
consequences in the other cases, one can just follow the Ref. \cite%
{ballistic}. We start by considering the existence of an electric field $F$
which is supposed to be directed along the arc $[-\theta _{0},\theta _{0}]$.
We must add the potential $eFr\theta $ in the Schrodinger equation (\ref%
{harmonic}). After expanding the effective potential in Eq. (\ref%
{harmonic}) up to the harmonic term around the minimum $x_{M}$, we rewrite
the Eq. (\ref{landau}) as
\begin{eqnarray}
E_{n} &=&\left( n+\frac{1}{2}\right) \Omega -eFr\arcsin {\left( \frac{\hbar
k_{Z}}{eBr}\right) }-\frac{e^{2}F^{2}}{2m\Omega ^{2}}  \notag \\
&-&\frac{\hbar ^{2}}{8mR^{2}}\left( 1+\frac{\beta ^{2}}{R^{2}}\right) ^{-1}+%
\frac{\hbar ^{2}}{m}\frac{\beta ^{2}}{a^{2}R^{2}},  \label{landau2}
\end{eqnarray}%
where
\begin{equation*}
\Omega \equiv \hbar \omega _{c}\left( 1+\frac{\beta ^{2}}{R^{2}}\right) ^{-%
\frac{1}{2}}\sqrt{1-\left( \frac{\hbar k_{Z}}{eBr}\right) ^{2}}.
\end{equation*}%
By defining $B^{\prime }\equiv B/\left( 1+\frac{\beta ^{2}}{R^{2}}%
\right) ^{\frac{1}{2}}$, we can rewrite the expression (\ref{landau2}) as
\begin{align}
E_{n}& =\left( n+\frac{1}{2}\right) \frac{\sqrt{1-\left( \frac{l_{B}^{\prime
2}k_{z}}{r}\right) ^{2}}}{ml_{B}^{^{\prime }2}}-eFr\arcsin {\left( \frac{%
l_{B}^{\prime 2}k_{z}}{r}\right) }  \notag \\
& -\frac{e^{2}F^{2}m^{2}l_{B}^{\prime 4}}{2m\left( 1-\frac{l_{B}^{\prime
4}k_{z}^{2}}{r^{2}}\right) }-\frac{\hbar ^{2}}{8mR^{2}}\left( 1+\frac{\beta
^{2}}{R^{2}}\right) ^{-1}+\frac{\hbar ^{2}}{m}\frac{\beta ^{2}}{a^{2}R^{2}}.
\label{landau3}
\end{align}%
We evoke the result of the Hall conductivity on a cylinder without any
defect which is found in Ref. \cite{ballistic}, namely
\begin{equation}
G_{H}=G_{0}\left[ 1-\frac{\arcsin {\left( \sin \phi _{0}-\pi \phi
_{0}N_{s}l_{B}\right) }}{\phi _{0}}\right],
\end{equation}%
which is valid for $\cos \phi _{0}>1/3$. In this expression, $G_{0}\equiv
2e^{2}/h$ is the conductance quantum, $l_{B}^{2}\equiv \hbar /eB$ and $%
N_{s}\equiv \left( m/2\pi \hbar ^{2}\right) E_{F}$ is the electronic
density. In the case with a screw dislocation, we make the changes $%
l_{B}^{2}\rightarrow l_{B}^{\prime 2}$, $\phi _{0}\rightarrow \theta
_{0}\equiv \phi _{0}+\beta z_{0}/\left( R^{2}\sqrt{1+\frac{\beta ^{2}}{R^{2}}%
}\right) $ and
\begin{align}
N_{s}& \rightarrow N_{s}^{\prime }  \notag \\
& \equiv \frac{m}{2\pi \hbar ^{2}}\left( E_{F}+\frac{\hbar ^{2}}{8mR^{2}}%
\left( 1+\frac{\beta ^{2}}{R^{2}}\right) ^{-1}-\frac{\hbar ^{2}}{m}\frac{%
\beta ^{2}}{a^{2}R^{2}}\right).
\end{align}%
Notice that the curvature potential and the contribution due to the
deformed potential modify the electronic density. This was observed in the
case of curvature only in Ref. \cite{saxena}.

\begin{figure}[h!]
\begin{center}
\includegraphics[height=7cm]{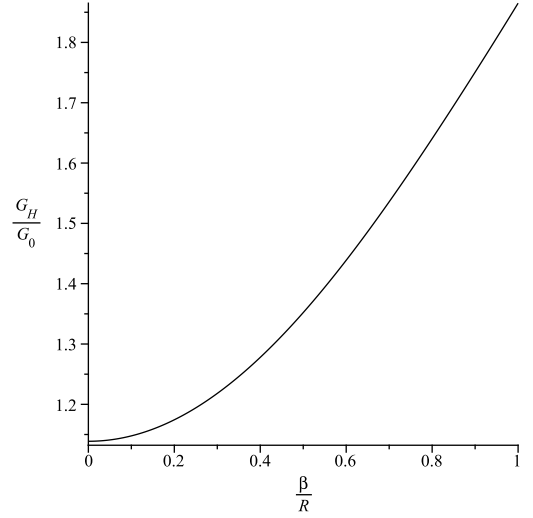}
\end{center}
\vspace{1.0cm}
\caption{\protect\scriptsize Enhancement of the Hall conductivity as $\frac{\beta}{R}$ is increased. We considered $l_B/R=0.5\;,$ $\phi_0=0.4\pi$ and $\frac{2mR^2 E_F}{\hbar^2}=2.$}
\label{fig8}
\end{figure}

For $\theta _{0}<<1$, we have
\begin{equation}
G_{H}=\frac{N_{s}^{\prime}e}{B^{\prime }}+\frac{G_{0}\theta _{0}^{2}}{%
12}\left[ 1-\left( 1-\frac{\nu^{\prime }}{2}\right) ^{3}\right],
\end{equation}%
Considering the expressions above and $z_{0}=b$, the Hall conductivity can
be ready as
\begin{align}
G_{H}& =\frac{\frac{m}{2\pi \hbar ^{2}}\left( E_{F}+\frac{\hbar ^{2}}{8mR^{2}%
}\left(1+\frac{\beta ^{2}}{R^{2}}\right) ^{-1}-\frac{\hbar ^{2}}{m}\frac{%
\beta ^{2}}{a^{2}R^{2}}\right) e}{B\left( 1+\frac{\beta ^{2}}{R^{2}}\right)
^{-\frac{1}{2}}}  \notag \\
& +\frac{G_{0}\left( \phi _{0}+2\pi \frac{\beta ^{2}}{R^{2}\sqrt{1+\frac{%
\beta ^{2}}{R^{2}}}}\right) ^{2}}{12}\left[ 1-\left( 1-\frac{\nu^{\prime }}{%
2}\right) ^{3}\right],
\end{align}%
where $\nu ^{\prime }\equiv 2\pi N_{s}^{^{\prime }}l_{B}^{{\prime}2}$. From Fig. \ref{fig8}, we can see that the screw dislocation enhances the hall conductivity as the Burges vector is increased.

\section{Concluding Remarks}

In conclusion, we studied the effect of a screw dislocation on the energy
levels of a 2DEG confined on a cylindrical surface. We have considered the
effects of two contributions: the covariant term which comes from the
geometric approach in the continuum limit and a noncovariant repulsive
scalar potential. Both appear due to elastic deformations on a semiconductor
with such kind of topological defect. This case, showing such
deformed potential can be thought in the context of 3D semiconductors where
the electrons are confined on a cylindrical shell. The absence of this
noncovariant potential could find place in the context of carbon nanotubes
which are intrinsically two dimensional entities and such potential has not
been derived. We have found that this noncovariant term changes
significantly the energy levels of electrons in this system. In the absence
of any magnetic field, the results can be explored in the context of quantum
rings. When a constant external magnetic field is present, we observed
significant modifications in the Landau levels and this will have important
consequences in the Quantum Hall effect. In fact, the diminished effective
frequency enhances the hall conductivity in the ultraquantum limit. In the other regimes, the steps of the Hall
conductivity usually shift to higher magnetic fields in this case\cite{anderson,andre}.
The quantum Hall effect on a cylinder without any defect was investigate in
reference \cite{hallcylinder} and it can be a start point if one has
intention to evaluate deeply the influences of a screw dislocation is this phenomenon.
Measurements of Hall conductivity in this particular geometry are
provided in Refs. \cite{hallexp1,hallexp2}. Again, thermodynamical properties in such geometry without defects are theoretically investigated in \cite{termo}.

As a final word, we remember the reader that we have investigated the 2DEG
in the harmonic approximation, which is valid for a sufficiently strong
normal to the surface component of the magnetic field and for electrons with
the Landau oscillator suspension center far enough from the side edges of a
cylindrical strip. If higher terms in the Hamiltonian is to be take into
account, the Ref. \cite{iran} can be used in order to treat the
problem via the perturbation theory. The authors have done this, again, for
a 2DEG on a cylinder without any kind of defect.

\section*{Acknowledgments}

This work was supported by the CNPq, Brazil, Grants No. 482015/2013-6 (Universal), No. 476267/2013-7 (Universal), No. 306068/2013-3 (PQ); FAPEMA, Brazil, Grant No. 00845/13 (Universal) and. project No. 01852/14 (PRONEM).

\end{document}